\documentclass[letterpaper, 10 pt, conference]{ieeeconf}  
\IEEEoverridecommandlockouts
\usepackage{amsmath,amssymb,amsfonts}
\usepackage{graphicx}
\usepackage{booktabs}
\usepackage{xcolor}
\usepackage{cite}
\usepackage{siunitx}
\usepackage{quantikz}

\usepackage{subcaption}
\usepackage{hyperref}

\DeclareMathOperator{\Acc}{Acc}

\newcommand{\ketbra}[2]{\ket{#1}\!\bra{#2}}

\usepackage[normalem]{ulem}
\usepackage{enumerate}

\title{\LARGE \bf 
Hardware-Agnostic Modeling of Quantum Side-Channel Leakage via Conditional Dynamics and Learning from Full Correlation Data
}

\author{%
Brennan Bell$^{1}$, Andreas Tr\"ugler$^{1,2}$, Konstantin Beyer$^{3}$, Paul Erker$^{4,5}$%
    \thanks{$^{1}$ Department of Data Privacy at Know Center GmbH, Graz, Austria.
        {\tt \small bell.brennan.p@gmail.com}}%
    \thanks{$^{2}$University of Graz, Graz, Austria.
        {\tt \small andreas.truegler@uni-graz.at}}%
    \thanks{$^{3}$Stevens Institute of Technology, Hoboken, NJ, USA.
        {\tt \small kbeyer1@stevens.edu}}%
    \thanks{$^{4}$Technische Universit\"at Wien, Vienna, Austria.}%
    \thanks{$^{5}$IQOQI Vienna, Austrian Academy of Sciences, Vienna, Austria.
        {\tt \small paul.erker@tuwien.ac.at}}%
}

\begin{document}

\maketitle
\thispagestyle{empty}
\pagestyle{empty}

\begin{abstract}
We study a sequential coherent side-channel model in which an adversarial probe qubit interacts with a target qubit during a hidden gate sequence. Repeating the same hidden sequence for $N$ shots yields an empirical \emph{full-correlation record}: the joint histogram $\widehat{P}_g(b)$ over probe bit-strings $b\in\{0,1\}^k$, which is a sufficient statistic for classical post-processing under identically and independently distributed (i.i.d.)\ shots but grows exponentially with circuit depth. We first describe this sequential probe framework in a coupling- and measurement-agnostic form, emphasizing the scaling of the observation space and why exact analytic distinguishability becomes intractable with circuit depth.

We then specialize to a representative instantiation (a controlled-rotation probe coupling with fixed projective readout and a commuting $R_x$ gate alphabet) where we (i) derive a depth-dependent leakage envelope whose maximizer predicts a coupling band as a function of depth {if the measurement data is reduced to marginal statistics}, and (ii) provide an operational decoder, via machine learning, a single parameter-conditioned map from $\widehat{P}_g$ to Alice's per-step gate labels, generalizing across coupling and noise settings without retraining. 
\end{abstract}


\section{Introduction}
We study a multi-tenant (multi-programmed) quantum processor setting where residual cross-talk creates a side-channel: while Alice executes a depth-$k$ circuit, a nearby adversarial probe qubit (Eve) weakly couples to Alice’s register and is repeatedly measured. Eve’s measurement statistics leak information about Alice’s gate sequence, even though Eve never directly measures Alice’s qubits. 

{It is well known that computing platforms}
can be attacked through e.g.\ timing and power side-channels, revealing secrets even when the algorithms themselves are cryptographically sound \cite{kocher1996timing, kocher1999differential}. At scale, micro-architectural leakage \cite{kocher2019spectre, lipp2018meltdown} and learning-based exploitation \cite{benadjila2018deep, maghrebi2020deep, picek2023sok, goodfellow2016deep} show that weak and noisy signals can become actionable after statistical amplification. 

Quantum platforms introduce an additional failure mode: \emph{coherent} leakage. An auxiliary degree of freedom can entangle with an otherwise isolated computation and acquire phase- and history-dependent signatures that are not well modeled as purely classical noise. This motivates sequential settings where information is transferred repeatedly and then partially erased by measurement back-action and open-system noise \cite{breuer2002open, schlosshauer2007decoherence}. Concretely, Alice applies an unknown length-$k$ gate string on a qubit $A$, while Eve couples a probe qubit $E$ after each gate and performs mid-circuit measurements on $E$. Figure~\ref{fig:protocol} summarizes the Alice/Eve interaction and Eve’s measurement pipeline. The resulting record is history-dependent: even for fixed coupling strength, statistics at step $t$ depend on the prior gate context through the evolving joint state.

Importantly, \emph{quantum-computing security} has already begun to document practical side-channel surfaces in today’s cloud and multi-user stacks. Timing observations in cloud quantum services can leak information about executed workloads and backend selection \cite{lu2025_timing_sca_qc}, while power/control-layer observables can reveal pulse-level structure sufficient for circuit reconstruction \cite{xu2023quantum_power_sca,erata2024_qc_reconstruction}. In multi-tenant regimes, co-located jobs can interact via device physics (e.g., crosstalk), creating additional attack surfaces \cite{choudhury2024_crosstalk_sca}. Our focus is complementary: we study a \emph{minimal coherent probe} mechanism that produces a correlation-rich mid-circuit record, and we ask when that record supports \emph{strict sequence recovery} under depth of the gate sequence, noise, and shot limits.

A key methodological choice in this paper is the observation model. 
Rather than restricting Eve to per-step marginals, we allow Eve to retain the \emph{full correlation structure over time}: repeating the \emph{same} hidden sequence of gates for $N$ shots yields an empirical joint histogram $\widehat{P}_g(b)$ over probe bit-strings $b\in\{0,1\}^k$. This correlation-rich record is the natural sufficient statistic for classical post-processing under i.i.d.\ shots, but it is high-dimensional ($2^k$).
{We analyze the case of data reduced to its per-time-step marginals as well as the full data with} explicit representational and inductive-bias choices for learning. 
Our threat model is motivated by the cross-talk and simultaneous-execution literature on shared quantum hardware, where imperfect isolation can correlate co-scheduled programs and leak information across tenants \cite{mehra2024defending_crosstalk, maurya_scv_readout_2024, choudhury2024_crosstalk_sca, harper_crosstalk_2025}.

\subsection{Contributions}

\textbf{General sequential-probe model and observation record.}
We formalize a sequential coupling-and-measurement threat model and adopt the full-correlation histogram $\widehat{P}_g(b)$ as Eve's sufficient statistic under i.i.d.\ shots, clarifying the exponential scaling of the observation space and the resulting limits on closed-form distinguishability beyond small depth.

\textbf{Instantiation-specific analytic predictor (controlled rotations + fixed readout).}
For a controlled-rotation probe coupling with fixed projective readout and a commuting $R_x$ gate alphabet, we derive a depth-dependent leakage envelope and a closed-form maximum of the envelope, $\theta^*(k)$ in Eq.~\eqref{eq:app_theta_star} for the coupling strength that maximizes distinguishability under repeated interactions, with the assumption that Eve only uses marginal statistics. This leakage envelope highlights where coupling/noise are neither so small that Eve sees nothing, nor so large that the side-channel becomes either trivial or washed out, yielding maximal learnability in practice.

\textbf{Amortized decoding from full correlations.}
Concretely, we train supervised sequence decoders (neural models) to predict gates from Eve’s observed statistics. We evaluate learnability both via information metrics (distribution distinguishability) and via trained decoders that attempt gate-by-gate reconstruction. The decoders are compact and parameter-conditioned, ingesting $\widehat{P}_g$ directly (not only marginals) and recovering Alice’s per-step gate labels across coupling and noise grids without retraining per grid point.

\section{Related Work and Positioning}
\textbf{Quantum side-channels in cloud stacks.}
Beyond classical side-channel analysis (SCA), recent work shows that quantum cloud services expose measurable side-channel surfaces tied to the control and execution stack. For example, timing observations on cloud-based quantum services can leak information about the executed workload and service behavior \cite{lu2025_timing_sca_qc}. In addition, power/control-layer side-channels can reveal pulse- and gate-level structure and allow reconstruction of quantum circuits executed on a controller \cite{xu2023quantum_power_sca,erata2024_qc_reconstruction}. These results motivate threat models where adversaries exploit \emph{classical} information that correlate with quantum computation.

\textbf{Multi-tenant, device-physics-mediated leakage and mitigations.}
A separate thread studies leakage that arises from shared-hardware physics, especially in multi-tenant (and similarly, in multi-programmed) settings, where co-located jobs may interact through mechanisms such as crosstalk \cite{choudhury2024_crosstalk_sca, harper_crosstalk_2025, almaguer2025rowhammer}. On the defense side, proposals that reshape effective interaction (e.g., through dynamical decoupling) aim to suppress crosstalk-mediated attacks and reduce information flow through coherent couplings \cite{mehra2024defending_crosstalk, harper_crosstalk_2025}. Classical ML-based side-channel defenses in non-quantum hardware provide a useful analogy for data-driven mitigation, but they operate on classical traces rather than coherent probe instruments \cite{yan_mlsca_2023}. Together, these lines underscore that quantum side-channel risk is not only a question of classical telemetry, but can also be rooted in coherent, hardware-level interactions.

\textbf{Information-theoretic side-channel analysis.}
Classical side-channel analysis often models leakage as a channel from a secret variable to an observed trace and studies distinguishability, mutual information, success rate, and trace complexity \cite{standaert_malkin_yung_2009, gierlichs_batina_tuyls_preneel_2008, decherisey_guilley_rioul_piantanida_2019, cheng_liu_guilley_rioul_2022}. Our setting follows the same inference-channel viewpoint after Eve's readout is fixed: the hidden gate sequence induces a classical observation law over probe records. The novelty is that this classical channel is generated by a coherent, history-dependent quantum interaction rather than by a conventional power, timing, or electromagnetic trace.

Before Eve's readout is fixed, the analogous quantum question is closer to state or channel discrimination, where optimal distinguishability is governed by quantum detection and channel-discrimination bounds \cite{helstrom_1976, audenaert_calsamiglia_2007, pirandola_laurenza_lupo_pereira_2019}. In this work, however, we fix Eve's measurement protocol and analyze the resulting classical observation channel.

\textbf{Our approach: coherent, correlation-rich leakage}
Unlike existing works that focus on classical telemetry (timing/power) or hardware-level crosstalk, our work studies estimator-independent constraints and operational learnability for sequence recovery when an adversary retains the full temporal correlation structure $\widehat{P}_g(b)$.

We focus on a minimal \emph{coherent probe} model: Eve gains information by repeatedly interacting a probe degree of freedom with the computation and measuring mid-circuit, producing a history-dependent record shaped by back-action and open-system contraction \cite{breuer2002open, schlosshauer2007decoherence, chitambar_resource_2019}. Our main departure from marginal-only analyses is to treat the full-correlation histogram $\widehat{P}_g(b)$ as the  observation primitive. This ensures that any observed failure of strict recovery can be attributed primarily to \emph{physics} (disturbance, contraction with depth, and finite-shot estimation), rather than to discarded temporal correlations. Thus, we treat cross-talk as an \emph{inference channel} and ask when it enables strict sequence recovery.

\begin{figure*}[htb!]
\centering
\begin{quantikz}[row sep=0.25cm, column sep=0.35cm] 
\lstick{$A$} & 
  \gate{g_1} & \gate[2]{V(\theta)} & \qw & \slice{} & 
  \gate{g_2} & \gate[2]{V(\theta)} & \qw & \slice{} & 
  \dots & \gate{g_k} & \gate[2]{V(\theta)} & \qw \\ 
\lstick{$E$} & 
  \qw &  & \meter{} \qw\wireoverride{n} & \slice{}\wireoverride{n} & 
  \qw &  & \meter{} \qw\wireoverride{n} & \slice{}\wireoverride{n} & 
  \dots & \qw &  & \meter{} \qw\wireoverride{n} 
\end{quantikz}
\vspace{0.5em}
\caption{Sequential coupling protocol: after each hidden gate $g_i$, Eve couples a probe through the gate \(V(\theta)\) and measures it mid-circuit, yielding a probe bit-string over depth $k$. Repeating $N$ times yields the histogram $\widehat{P}_g(b)$.}
\label{fig:protocol}
\end{figure*}

\section{Model and Threat Setting}

\subsection{General sequential probe model}
Alice's hidden gate sequence is $g_{1:k}\in\mathcal{G}^k$ with a known alphabet $\mathcal{G}=\{U^{(1)},\dots,U^{(M)}\}$ acting on a target qubit $A$. At each step $t$, Alice applies $U^{(g_t)}$ on $A$, then $A$ interacts with Eve's probe $E$ via a fixed two-qubit unitary $V(\theta)$, where \(\theta\) is a coupling strength, followed by a dichotomic measurement on $E$ described by a POVM $\{M_{y_t}^{(t)}\}$. 
The choice of the measurement can in principle vary from step to step and may depend on previous outcomes.
Noise is modeled by a CPTP map \(\mathcal{N_\lambda}\), where \(\lambda\) determines the strength, interleaved at each step. This is naturally described as a sequential quantum instrument: each step combines Alice's gate, Eve's coupling, Eve's measurement, and the intervening noise channel into an outcome-indexed completely positive map \cite{davies_lewis_1970, nielsen_chuang_2010, watrous_2018}. One execution yields a probe bit-string $b=(y_1,\dots,y_k)\in\{0,1\}^k$ drawn from the underlying probability distribution $P_g(b)$ induced by $(V,\{M_y\},\mathcal{N}_\lambda)$.
The structure of the scenario is shown in Fig.~\ref{fig:protocol}.

\subsection{What is (and is not) "hardware-agnostic"}
We do not assume a microscopic Hamiltonian for the coupling; instead we treat $V(\theta)$  and the measurement $\{M_y\}$ as effective primitives summarizing the interaction and readout in a given protocol. However, learnability \emph{does} depend on the choice of coupling and measurement; changing just Eve's measurement basis generally changes the induced laws and can materially alter strict recovery.

The only invariances we rely on are consistent local re-parameterizations that preserve the induced statistics. For example, for any probe unitary $W$,
\begin{equation}
    V \mapsto (I\otimes W)\,V\,(I\otimes W^\dagger),\qquad
    M_y \mapsto W\,M_y\,W^\dagger,
\end{equation}
leaves $P_g(b)$ unchanged since it corresponds to a basis change on Eve's probe applied consistently to both the coupling and the measurement. Analogous joint conjugations on Alice's side (conjugating both the coupling and the gate alphabet) are statistically equivalent. Outside such consistent transformations, different couplings/measurements correspond to different threat models.

\subsection{Full-correlation observation model}
One circuit execution yields a probe bit-string $b=(y_1,\dots,y_k)\in\{0,1\}^k$. Repeating the \emph{same} hidden sequence for $N$ shots produces counts
\begin{equation}
    C(b)=\sum_{i=1}^N \mathbf{1}\{b^{(i)}=b\},\qquad
    \widehat{P}_g(b)=C(b)/N,
\end{equation}
an empirical distribution on $\{0,1\}^k$ (dimension $2^k$). This is the full-correlation record. 
Under i.i.d.\ repetitions conditioned on the same hidden sequence, \(C\) is a multinomial count vector with parameter \(P_g\), so \(\widehat P_g\) is the empirical law of Eve's trajectory distribution \cite{CoverThomas2006}.

\subsection{Goal and metrics}
Eve aims to infer $g_{1:k}$ from the empirical distribution $\widehat{P}_g$ and auxiliary conditioning parameters (coupling $\theta$ and noise scale $\lambda$). We evaluate:
(i) per-position accuracy
at step \(t\)
\begin{equation}
    \Acc_{\mathrm{position}}(n)
    =
    \Pr\!\left[\hat g_{t}=g_{t}\right],
\end{equation}
as well as (ii) \emph{strict sequence accuracy} for sequences of length $k$,
which is given by
\begin{equation}
    \Acc_{\mathrm{strict}}(k)
    =
    \Pr\!\left[\hat g_{1:k}=g_{1:k}\right].
\end{equation}

\section{Intractability in pairwise distinguishability}
\label{sec:intractability-roots}

A decoder can succeed only if different hidden gates or gate sequences induce distinguishable observation laws on Eve's record. For a full sequence \(g\in\mathcal G^k\), let
\begin{equation}
    P_g(b),\qquad b\in\{0,1\}^k,
\end{equation}
denote the infinite-shot distribution of Eve's probe bitstring. Finite-shot data \(\widehat P_g\) are empirical estimates of this law.

There are two related discrimination problems. For strict sequence recovery, the classes are the individual sequence-conditioned laws \(P_g\). For position recovery at step \(t\), the classes are the gate-conditioned mixture laws
\begin{equation}
    \bar P_{t,a}(b)
    =
    \mathbb E\!\left[P_g(b)\mid g_t=a\right],
    \qquad a\in\mathcal G .
\label{eq:position_class_law}
\end{equation}
Thus position recovery asks whether the three mixture laws \(\{\bar P_{t,a}\}_{a\in\mathcal G}\) are distinguishable, while strict recovery asks whether the full family \(\{P_g\}_{g\in\mathcal G^k}\) is distinguishable.

To express this estimator-independent question, one may use standard statistical distances or divergences such as total variation, Kullback--Leibler divergence, or Jensen--Shannon divergence \cite{CoverThomas2006, lin1991}. If a geometry on the histogram bins is specified, there exist computable proxies for distributional separation \cite{cuturi_sinkhorn_2013}. Inequalities such as Pinsker's relate divergence bounds to total-variation separation \cite{schutzenberger_thesis_1954, pinsker_info_1964}. For example, for a generic distinguishability functional \(D\), define
\begin{equation}
    D_{\mathrm{pos}}^{(t)}(a,a')
    = 
    D\!\left(\bar P_{t,a},\bar P_{t,a'}\right),
    \qquad a\neq a',
\label{eq:D_position}
\end{equation}
for position recovery, and
\begin{equation}
    D_{\mathrm{seq}}(g,g')
    =
    D\!\left(P_g,P_{g'}\right),
    \qquad g\neq g',
\label{eq:D_sequence}
\end{equation}
for strict sequence recovery. If these distances vanish for two classes, then those classes are information-theoretically indistinguishable from Eve's record, which is a decoder-independent observation.

The accuracy heatmaps Figs.~\ref{fig:position-accuracy}--\ref{fig:comparison} do not directly evaluate Eqs.~\eqref{eq:D_position}--\eqref{eq:D_sequence}; instead, they report operational decoder accuracy. The distances are used here only to clarify the underlying statistical discrimination problem, and to explain why, in regimes where the induced observation laws are nearly identical, no decoder should be expected to recover the hidden gates reliably.

\subsection{Why analytic distinguishability becomes intractable}
The object $\bar P_{t,a}$ is an expectation over all length-$k$ continuations consistent with the sequence $u$. Even in the noiseless case, writing it in closed form entails composing $k$ sequential instruments (unitary coupling + measurement) and averaging over $|\mathcal{G}|^{k-d}$ continuations, while each $P_g$ itself is a distribution on $2^k$ outcomes. Thus, the exact evaluation of Eqs.~\eqref{eq:D_position}--\eqref{eq:D_sequence} scales exponentially in both depth and outcome dimension, and symbolic expressions rapidly become unwieldy beyond very small $k$. This motivates (i) marginals as proxies that track the dominant informative mode, and (ii) amortized decoders trained directly on $\widehat P_g$.

Finite-shot statistics add a second combinatorial layer. For the full-correlation record, \(N\) repetitions produce a multinomial histogram over \(2^k\) bitstrings. Exact summation over all possible histograms therefore involves
\begin{equation}
    \binom{N+2^k-1}{2^k-1}
\end{equation}
multinomial count types. If Eve retains only per-time marginal counts \(K=(K_1,\ldots,K_k)\), the corresponding count space has
\begin{equation}
    (N+1)^k
\end{equation}
possible values. These standard count-space scalings are the reason that exact finite-shot distinguishability is feasible only at small depth or small shot count \cite{CoverThomas2006}.

\section{A proof-of-principle example}
In order to illustrate important features of such a side-channel model, we now turn to a concrete example. 

\begin{enumerate}[(i)]
\item The coupling gate between the qubits is assumed to be a controlled \(X\) rotation with Eve's qubit being the target.
\begin{align}
    V(\theta) = \ketbra{0}{0}\otimes I + \ketbra{1}{1}\otimes R_x(\theta).
\end{align}
This gate is entangling for all \(\theta\mod 2\pi \neq 0\) and can therefore leak information from Alice's to Eve's system.

\begin{figure*}[ht!]
\centering
\includegraphics[width=\linewidth]{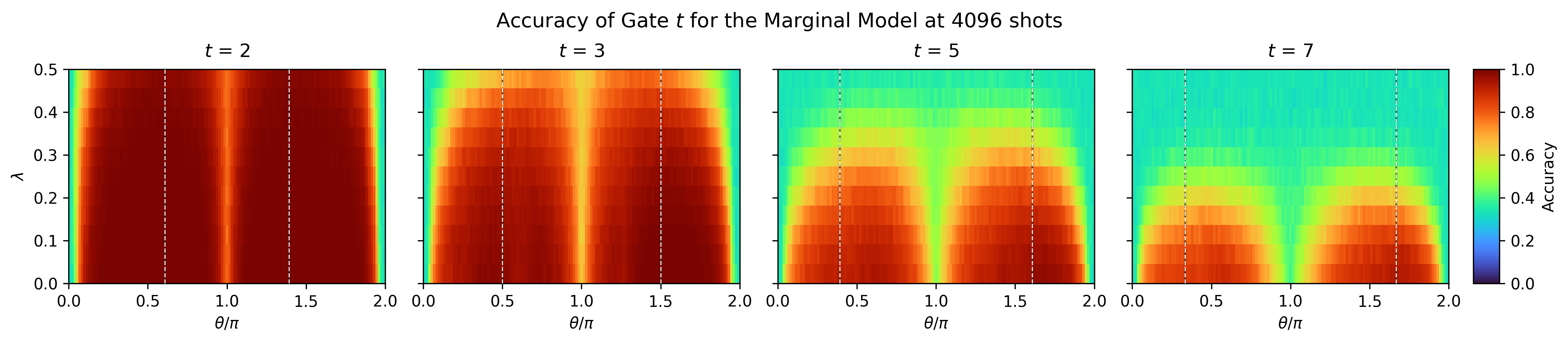}
\caption{
Accuracy for correct detection of the gate at position \(t\) as a function of coupling strength \(\theta\) and noise parameter \(\lambda\), using only marginal measurement statistics and discarding correlations between Eve's measurement outcomes. The dashed vertical lines correspond to the predicted accuracy maxima for \(\lambda = 0\), \(\theta^*(t)\) and \(2\pi-\theta^*(t)\), where \(\theta^*(t)=2\arcsin\sqrt{2/(t+1)}\), as derived in Eq.~\eqref{eq:app_theta_star}. Increasing \(t\) reduces the available marginal signal because earlier probe interactions contract the state entering the \(t\)-th gate. Results are shown for 4096 shots.
}
\label{fig:position-accuracy}
\end{figure*}

\item Eve performs a fixed projective $Z$-measurement on $E$ after each interaction.

\item Eve reinitializes her probe to a fresh state $\ket{0}$ before interacting with Alice’s register.

\item Noise on \(A\) is modeled as gate-independent depolarizing noise applied after each timestep: 
\begin{align}
\label{eq:noise-channel}
    \mathcal{N}_\lambda[\rho] = (1-\lambda)\rho+\lambda\,\mathbb{I}/2.  
\end{align}
This noise model is used as a simple heuristic contraction mechanism for the side-channel inference problem. It should be distinguished from error-mitigation or implementability frameworks, which ask how target operations can be represented or recovered in the presence of noise \cite{jin_noisy_2025}.

\item The gate alphabet is the commuting set $\mathcal{G}=\{R_x(\pi/8),R_x(\pi/2),R_x(\pi)\}$.
\end{enumerate}

In the following we will analyze this exemplary case with respect to two different decoding strategies for Eve. First, we will restrict the available data to the marginal probability at each time step \(t\), i.e., neglecting the correlations in Eve's measurement data. This approach avoids an exponential growth of complexity in the data but introduces unavoidable quantum noise. We derive an optimal coupling strength to minimize this influence.

Second, we investigate Eve's ability to reconstruct Alice's gate sequence from the full measurement statistics with machine learning methods. Of particular interest is the finite shot regime. We find that Eve can reach high accuracies even for sparse measurement data.

\section{Learning from marginal distribution}

\label{sec:marginal}

According to the fundamental quantum principle of no information without disturbance, the leakage from \(A\) to \(E\) introduces noise in Alice's system, even for an otherwise noiseless scenario with \(\lambda = 0\) in Eq.~\eqref{eq:noise-channel}.
This noise becomes relevant if Eve disregards the correlation in her measurement data. 
Alice's average state at time step \(t\) is mixed due to the previous interactions with Eve's system. Thus, Eve's marginal distribution at \(t\) inherits this noise.
(Note that this is not the case if Eve keeps track of the correlations in her data because conditioned her measurement record, Alice's system is always in a pure state if \(\lambda = 0\).)

The stronger the coupling \(\theta\), the stronger the noise. On the other hand, a stronger coupling increases the measurement signal on Eve's side. This interplay leads to a characteristic coupling regime which we show in Fig.~\ref{fig:position-accuracy}. The accuracy of guessing gate \(t\) correctly is plotted for \(t \in \{2,3,5,7\}\) as a function of the coupling strength \(\theta\) and the noise parameter \(\lambda\). As expected, the accuracy drops for increasing \(t\). Furthermore, the ideal coupling strength, which maximizes the accuracy, shifts slightly.

For the noiseless case \(\lambda = 0\), the ideal coupling strength can be predicted analytically.
For the controlled-rotation  used here, a useful analytic proxy is the gate-conditioned spread in a canonical per-step event probability, which can be written (up to an alphabet-dependent scale $\alpha_{\mathcal{G}}$) as
\begin{equation}
    \Delta p_t(\theta;\mathcal{G}) \approx \alpha_{\mathcal{G}}
    \sin^2\!\Bigl(\frac{\theta}{2}\Bigr)\cos^{t-1}\!\Bigl(\frac{\theta}{2}\Bigr).
\label{eq:delta-envelope}
\end{equation}
The $\sin^2(\theta/2)$ factor captures weak-coupling growth of transfer of information into the Eve's system, while the $\cos^{t-1}(\theta/2)$ factor captures the \((t-1)\)-fold introduction of noise to Alice's system \(A\) due to its coupling with \(E\) (see the Appendix for details).
Maximizing $f_t(\theta)=\sin^2(\theta/2)\cos^{t-1}(\theta/2)$ yields
\begin{equation}
    \theta^{*}(t)=2\arcsin\sqrt{\frac{2}{t+1}},
\label{eq:theta-star}
\end{equation}
which shifts to smaller coupling with depth. Operationally, $\theta^*(k)$ predicts where strict recovery should concentrate in coupling sweeps if Eve uses only per-step marginals of the full distribution.
We show \(\theta^*\) as dashed lines in Fig.~\ref{fig:position-accuracy}. Note that the optimal \(\theta\) shifts further under noise, i.e.\ \(\lambda > 0\).  

The accuracy of determining the full sequence of gates correctly is necessarily smaller than the single gate accuracy. We show it for circuit depth \(k=7\) and a varying number of shots in the finite statistics in Fig.~\ref{fig:comparison}\,(a). These plots based on the marginal statistics serve as the baseline comparison for the case of full statistics, including correlations between Eve's measurement outcomes, discussed in the next section.

\section{Learning from Full Correlation Data}
A recovery of the Alice's gate sequence based on marginal data only is constraint by construction due to unavoidable quantum noise, as motivated in the previous section. 
Here, we analyze the case where Eve uses the full measurement data including the correlations between sequential outcomes.

We learn a single, two-parameter-conditioned decoder
\begin{equation}
    f_\omega:\bigl(\widehat{P}_g,\theta,\lambda\bigr)\mapsto \hat g_{1:k},
\end{equation}
trained across a grid of $(\theta,\lambda)$ so that inference is amortized across physical regimes.

\subsection{Histogram-conditioned temporal regressor (Hist-TCN)}
Let $x\in\mathbb{R}^{2^k}$ denote the vectorized full-correlation histogram $x=\widehat{P}_g(\cdot)$ under a fixed bit-string ordering. We convert $x$ into a length-$k$ feature sequence by repeating it at each time step and appending a positional code and physical conditioning:
\begin{align}
    X_t = \bigl[\; \theta/\pi,\;\lambda,\; e_t,\; x \;\bigr]\in\mathbb{R}^{2+k+2^k},&& t=1,\dots,k,
\end{align}
where $e_t\in\{0,1\}^k$ is the one-hot vector indicating the position $t$. This yields $X\in\mathbb{R}^{k\times(2+k+2^k)}$, which is processed by a dilated temporal convolutional network (TCN) \cite{waibel_tdnn_1989, lea_tcn_2016, bai_convolutional_2018, yazdanbakhsh_timeseries_2019, kakuba_tdnn_2024} over the step index $t$ with a per-step regression head.

The network outputs a scalar $\hat a_t\in[0,1]$ interpreted as a normalized rotation angle $\hat\varphi_t/\pi$ at step $t$. Gate labels are obtained by nearest-neighbor decoding to the known alphabet $\{1/8,\,1/2,\,1\}$:
\begin{equation}
    \hat g_t = \arg\min_{m\in\{1/8,1/2,1\}} \bigl|\hat a_t - m\bigr|.
\end{equation}

We train with a smooth $\ell_1$ loss on the continuous targets $a_t\in\{1/8,1/2,1\}$. Known in the literature as the \emph{Huber} loss \cite{huber_robust_1964}, the function transitions from quadratic to linear growth beyond a threshold, improving robustness to heavy-tailed errors. This regression formulation is compact and stable under finite-shot noise, and it reuses a single decoder across the full $(\theta,\lambda)$ grid without per-point retraining \cite{girshick_fastrnn_2015}.

 We remark that we have also explored Gray-code reorderings \cite{gray_pulse_1947} and Walsh--Hadamard \cite{walsh_wht_1923} features for alternative decoders. Here we report Hist-TCN as the compact baseline used for the displayed full-histogram results; broader representation comparisons are left for future work.

\subsection{Why representation matters as dimension scales exponentially}

The full-correlation record is a distribution on \(\{0,1\}^k\). After vectorization, its \(2^k\) bins inherit an ordering that is not physically unique; different orderings impose different artificial notions of locality on the input. This matters especially for convolutional architectures, since a TCN uses local filters over its input representation and over the sequence index. Although \(k=7\) gives only \(128\) histogram bins in the experiments reported here, the dimension doubles with each additional probe measurement, so representation choices become increasingly important at larger depths.

In the reported full-histogram experiments, we use the vectorized histogram directly as part of the per-position TCN input, together with the physical parameters \((\theta,\lambda)\). Natural extensions include orderings or transforms that expose additional structure with simple inductive biases: Gray-code ordering would make neighboring histogram indices differ by one bit, which may be useful for local convolution, while the Walsh--Hadamard transform would map the histogram to parity/correlation spectra, which may be useful for models with global receptive fields.

\subsection{Conditioning on physical parameters}
We concatenate normalized $(\theta,\lambda)$ to the learned embedding before the prediction head. This yields a single amortized decoder that interpolates across the full grid rather than retraining a model per coupling/noise setting.

\begin{figure*}[htb!]
\centering
\includegraphics[width=\linewidth]{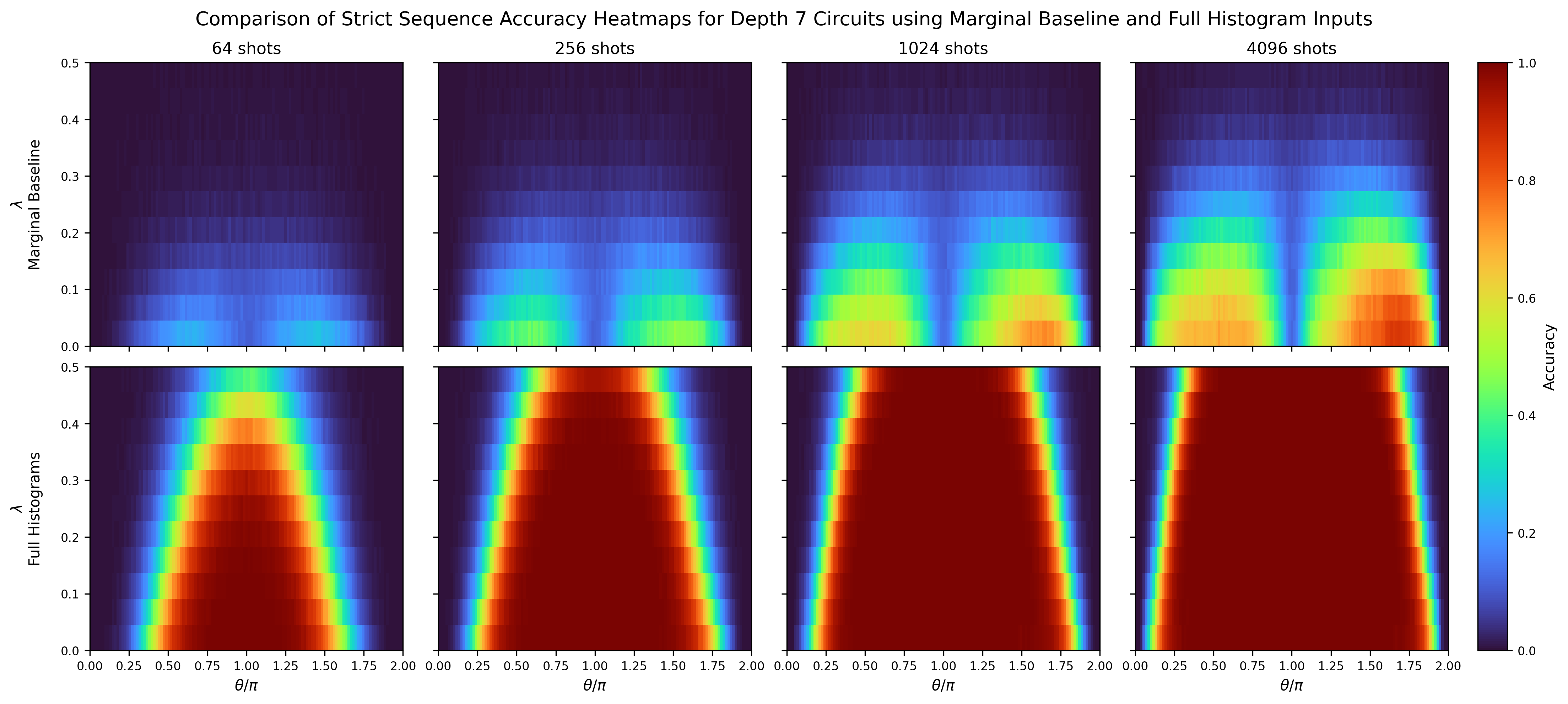}
\caption{Accuracy for correct detection of the full gate sequence of depth \(k=7\) over coupling $\theta$ and depolarizing noise $\lambda$, shown across increasing shots $N$.
Subfigure (a) uses the marginal statistics at each time step \(t\) as the input. Subfigure (b) was obtained from the full statistics, including correlations between Eve's measurement outcomes. The optimal coupling strength for maximal accuracy differs considerably between the two cases.
In both cases, noise (y-axis) lowers accuracy and narrows the viable coupling strength as $\lambda$ increases, while finite-shot effects (indicated by $N$) represent information density improvements in Eve's data.}
\label{fig:comparison}
\end{figure*}

\subsection{Estimator-independent limit}

The trained decoder is only one estimator for the induced observation channel. Information-theoretic bounds give estimator-independent limits: if two gate sequences induce nearly identical observation laws, no decoder can reliably separate them. This is the sequence-recovery analogue of classical side-channel analyses that relate leakage channels to mutual information, success probability, and trace complexity \cite{standaert_malkin_yung_2009, gierlichs_batina_tuyls_preneel_2008, decherisey_guilley_rioul_piantanida_2019, cheng_liu_guilley_rioul_2022}.

Concretely, total variation and KL divergence control statistical distinguishability, while Pinsker-, Le~Cam-, and Fano-type inequalities relate these divergences to unavoidable error probabilities \cite{CoverThomas2006, Yu1997, Tsybakov2009, pinsker_info_1964}. In our model, repeated coupling and noise contract distinguishability with depth, while finite-shot sampling further blurs the empirical record. Thus low accuracy outside the high-recovery regions should not be interpreted only as a model failure; in those regions, the induced observation laws themselves may be too close to support strict recovery.

\section{Experimental Setup}
\subsection{Simulation}
We simulate two-qubit density-matrix dynamics with interleaved noise and mid-circuit measurement. A typical configuration:
\begin{itemize}
    \item Gate alphabet $\mathcal{G}=\{\mathrm{Rx}(\pi/8),\mathrm{Rx}(\pi/2),\mathrm{Rx}(\pi)\}$.
    \item Depth $k$ (e.g., $k=7$).
    \item Shots per sequence $N$ (sweep: e.g., $N\in\{32,64,\dots,4096\}$).
    \item Coupling grid $\theta\in[0,2\pi]$; noise grid $\lambda$ mapped to a depolarizing rate.
\end{itemize}
\medskip

We use the commuting \(R_x\) alphabet as a controlled proof-of-principle case: it isolates repeated probe coupling, mid-circuit measurement back-action, Markov noise, and finite-shot sampling without adding non-commutative control complexity. The side-channel mechanism itself does not require commutativity; changing the alphabet, coupling, or readout changes the induced observation laws and is expected to change the detailed learnability landscape.

\subsection{Training and evaluation}

Training data are generated by simulating the joint Alice--Eve dynamics for random gate sequences \(g\), producing either marginal finite-shot statistics or full finite-shot histograms over Eve bitstrings. The marginal baseline uses the per-time count statistics obtained from the level-0 marginal probabilities, whereas the full-histogram model uses the empirical joint record \(\widehat P_g\). We train parameter-conditioned neural decoders over the \((\theta,\lambda)\) grid and evaluate two metrics: per-position gate accuracy, shown for marginal statistics in Fig.~\ref{fig:position-accuracy}, and strict depth-\(7\) sequence accuracy, shown for marginal and full-histogram inputs in Fig.~\ref{fig:comparison}.

\section{Results}

Figure~\ref{fig:position-accuracy} shows the marginal model's per-position accuracy at \(N=4096\) shots for \(t\in\{2,3,5,7\}\). The marginal record is already highly informative for early positions, especially at low noise and intermediate-to-strong coupling. However, the recoverable region shrinks and weakens with position: by \(t=7\), high accuracy is confined to a smaller low-noise region. This is consistent with the non-selective back-action picture in outlined in the Appendix: when temporal correlations are discarded, successive probe interactions contract the Alice components that carry gate-dependent information into later marginal readouts.


The position-accuracy panels show that, on the principal branch \(\theta\in[0,\pi]\), marginal learnability is organized by the same signal--survival tradeoff captured by the envelope in Eq.~\eqref{eq:delta-envelope}. Very weak coupling gives little signal, while the neighborhood of \(\theta=\pi\) can produce a narrow, low-accuracy trough where the marginal class statistics become less distinguishable. Away from this symmetry-induced trough, broad, high-accuracy ranges persist at low noise, especially for shallow gate positions. The dashed \(\theta^*\) lines should therefore be interpreted as theoretical position-accuracy maxima for the marginal decoder, rather than as predictions of the sequence-accuracy maxima. The observed landscape reflects a mixture of local readout strength, back-action-induced contraction, and contextual information from earlier marginal counts.

Figure~\ref{fig:comparison} compares strict depth-\(7\) sequence recovery using marginal statistics and full histograms. The marginal baseline in Fig.~\ref{fig:comparison}\,(a) remains substantially below the full-histogram model and exhibits a relatively narrow recoverable region even at \(4096\) shots. In contrast, Fig.~\ref{fig:comparison}\,(b) shows that retaining the full joint histogram dramatically improves strict recovery: increasing \(N\) from \(64\) to \(4096\) expands a broad high-accuracy region over \(\theta\), with the strongest performance at low noise.

Noise suppresses both observation models. Moving upward in \(\lambda\) lowers accuracy and narrows the high-recovery region, as expected from additional depolarizing contraction on Alice's state. The effect is especially visible in strict sequence recovery, where all positions must be decoded correctly and small per-step errors compound across the full depth-\(7\) sequence.

Together, the two figures separate two effects. First, marginal statistics already contain gate information, but this information becomes increasingly fragile with depth and noise. Second, temporal correlations carry substantial additional information: retaining the full histogram converts the same underlying probe record into a much stronger strict-sequence decoder input, even at a small number of shots.

\section{Conclusion}

Operationally, Eve's side-channel yields gate-level information in a strongly parameter-dependent way. With only marginal statistics, Eve can recover early gates accurately in favorable coupling and low-noise regimes, but per-position accuracy degrades with depth and strict sequence recovery remains limited. This is the practical cost of discarding temporal correlations in the probe record.

Retaining the full-correlation histogram changes the picture substantially. For the same depth-\(7\) task and shot budgets, full-histogram inputs produce a much broader and higher-accuracy strict recovery region than marginal inputs. Thus the side-channel risk is not captured by a single leakage scalar or by per-time marginals alone; it depends jointly on coupling strength, noise, shot count, depth, and the representation of Eve's measurement record. These results support the view that coherent crosstalk should be analyzed as an inference channel over full measurement histories, with marginal envelopes serving as useful but incomplete interpretive guides.

\section*{Acknowledgments}
An earlier version of this work was presented at the 3rd annual Quantum Computer Cybersecurity Symposium in October 2025; we thank Jakub Szefer for the invitation and further thank the many participants at the symposium for their insights and feedback. BB, AT and PE acknowledge support by the Austrian Federal Ministry of Education, Science, and Research via the Austrian Research Promotion Agency (FFG) through the Quantum Austria project QUICHE (No.\ 914033). KB acknowledges support by the NSF under award No.~2239498, NASA under award No.~80NSSC25K7051 and the Sloan Foundation under award No.~G-2023-21102. This project is co-funded by the European Union (Quantum Flagship project ASPECTS, Grant Agreement No. 101080167). 

\appendix

We derive Eq.~\eqref{eq:delta-envelope} for the controlled-rotation example under the \emph{marginal-only} observation model. We assume that Eve re-initializes \(E\) in \(\ket{0}\), measures \(E\) in the \(Z\)-basis after each interaction, and discards temporal correlations. The interaction is
\begin{equation}
U_\theta=\ketbra{0}{0}_A\otimes I_E+\ketbra{1}{1}_A\otimes R_x(\theta)_E,\qquad
R_x(\theta)=cI-isX,
\label{eq:app_crx_reset}
\end{equation}
where \(c=\cos(\theta/2)\) and \(s=\sin(\theta/2)\). The corresponding Kraus operators on Alice are
\begin{equation}
K_0=\ketbra{0}{0}+c\ketbra{1}{1},\qquad
K_1=-is\ketbra{1}{1}.
\label{eq:app_kraus}
\end{equation}
For \(\rho_A=\frac12(I+\vec r\cdot\vec\sigma)\), Eve's probability of outcome \(1\) is
\begin{equation}
p_1(\rho_A,\theta)=\operatorname{Tr}(K_1^\dagger K_1\rho_A)
=s^2\rho_{11}
=\frac{1-r_z}{2}\sin^2\!\Bigl(\frac{\theta}{2}\Bigr).
\label{eq:app_probe_one_probability}
\end{equation}
Thus the marginal signal contains a transfer factor \(\sin^2(\theta/2)\), multiplied by the spread of the relevant \(r_z\)-component.

When Eve retains only per-time marginals, Alice evolves on average under the non-selective channel
\begin{equation}
\mathcal E_\theta(\rho)=K_0\rho K_0^\dagger+K_1\rho K_1^\dagger
\label{eq:app_nonselective_channel},
\end{equation}
which leads to a dephasing around the \(z\)-axis given by the Bloch contraction \((r_x,r_y,r_z)\mapsto(c r_x,c r_y,r_z)\), so each interaction damps coherence-like components by \(|\cos(\theta/2)|\). Depolarizing noise \(\mathcal{N}_\lambda\) in Eq.~\eqref{eq:noise-channel} adds an approximately \(\theta\)-independent Bloch contraction and therefore changes the envelope scale but not its maximizer in \(\theta\).

For the commuting \(R_x\)-alphabet, gate-conditioned contrasts visible in Eve's marginal record acquire approximately one factor of \(\cos(\theta/2)\) per step after discarding the previous measurement outcomes. Absorbing alphabet-dependent constants into \(\alpha_{\mathcal G}\), this gives
\begin{equation}
\Delta p_t(\theta;\mathcal G)\approx
\alpha_{\mathcal G}\sin^2\!\Bigl(\frac{\theta}{2}\Bigr)\cos^{t-1}\!\Bigl(\frac{\theta}{2}\Bigr),
\qquad 0\le\theta\le\pi,
\label{eq:app_delta_envelope}
\end{equation}
with the mirrored branch given by \(\theta\mapsto2\pi-\theta\). Finally, optimizing the one-parameter envelope \(f_t(\theta)=\sin^2(\theta/2)\cos^{t-1}(\theta/2)\) on the principal branch \(0\le\theta\le\pi\) gives
\begin{equation}
\theta^*(t)=2\arcsin\sqrt{\frac{2}{t+1}}.
\label{eq:app_theta_star}
\end{equation}
On the full coupling interval \(0\le\theta\le2\pi\), the envelope is symmetric under \(\theta\mapsto2\pi-\theta\), so the corresponding mirrored maximizer is \(2\pi-\theta^*(k)\). This maximum is used as an analytic reference for interpreting the accuracy of the noiseless marginal model and is specific to the prediction of a single gate for the controlled-rotation instantiation used in this paper; it is used in the marginal experiments only as a reference point.

\bibliographystyle{IEEEtran}
\bibliography{manuscript-references-redlined}
\end{document}